\documentclass[pra,a4paper,showpacs,byrevtex,preprint]{revtex4-1}
\usepackage{graphicx}

\usepackage{amsmath}
\usepackage{array}
\usepackage{bm}
\usepackage{amssymb}
\usepackage{amsfonts}

\begin{document}

\title{The quantum $N$-body problem with a minimal length}

\author{Fabien \surname{Buisseret}}
\thanks{F.R.S.-FNRS Postdoctoral Researcher}
\email[E-mail: ]{fabien.buisseret@umons.ac.be}

\affiliation{Service de Physique Nucl\'{e}aire et Subnucl\'{e}aire,
Universit\'{e} de Mons--UMONS,
Acad\'{e}mie universitaire Wallonie-Bruxelles,
Place du Parc 20, B-7000 Mons, Belgium}

\date{\today}

\begin{abstract}
The quantum $N$-body problem is studied in the context of nonrelativistic quantum mechanics with a one-dimensional deformed Heisenberg algebra of the form $[\hat x,\hat p]=i(1+\beta \hat p^2)$, leading to the existence of a minimal observable length $\sqrt\beta$. For a generic pairwise interaction potential, analytical formulas are obtained that allow to estimate the ground-state energy of the $N$-body system by finding the ground-state energy of a corresponding two-body problem. It is first shown that, in the harmonic oscillator case, the $\beta$-dependent term grows faster with $N$ than the $\beta$-independent one. Then, it is argued that such a behavior should be observed also with generic potentials and for $D$-dimensional systems. In consequence, quantum $N$-body bound states might be interesting places to look at nontrivial manifestations of a minimal length since, the more particles are present, the more the system deviates from standard quantum mechanical predictions. 
\end{abstract}



\maketitle

\section{Introduction}

The existence of a minimal observable length in Nature is an appealing suggestion of string theory and quantum gravity, see \textit{e.g.}~\cite{s1,s2,s3,s4,s5,s6}. Therefore, and also because of their intrinsic interest, the study of quantum theories characterized by a minimal length has become an active area in theoretical physics. An economical way of introducing such a minimal length is to modify the canonical commutation relations between the position and momentum operators in quantum mechanics, \textit{i.e.} to use a modified Heisenberg algebra~\cite{k0a,k0b,k0c}. As discussed in detail in these last references, in one dimension, an algebra of the form (in units where $\hbar=c=1$)
\begin{equation}\label{alg1}
	\left[\hat x,\hat p\right]=i\, \Theta(\hat p)
\end{equation}
is able to yield a minimal uncertainty on $\hat x$. The function $\Theta(\hat p)$ can be expanded in powers of $\hat p$. Assuming an isotropic situation and demanding to recover the standard Heisenberg algebra at the lowest order, one has at order $\hat p^2$
\begin{equation}\label{p2}
\Theta(\hat p)=1+\beta \hat p^2	.
\end{equation}

The ansatz~(\ref{p2}) is the simplest way of generating a minimal length. Indeed, the uncertainty relation  
\begin{equation}\label{eq2}
	\Delta\hat x\geq\frac{1}{2}\left(\frac{1}{\Delta \hat p}+\beta\Delta\hat p\right)
\end{equation}
imposes a nonzero minimal uncertainty on $ \Delta\hat x$, given by $\sqrt\beta$. One- or two-body problems have been well studied using the modified algebra defined by Eqs.~(\ref{alg1})-(\ref{p2}), especially the harmonic oscillator~\cite{minic,dadic,quesne1,quesne2}, the hydrogen atom~\cite{Brau1,ben,mazi,bouaz}, and the gravitational quantum well~\cite{brau2,bane}. Remark that the parameter $\beta$ should be such that $\beta \langle\hat p^2\rangle \ll1$, otherwise such a modification would have been already detected experimentally. The most stringent upper bound on the minimal length scale obtained so far is the one coming from the hydrogen atom and is equal to $3.3\, 10^{-18}$ m, leading to $\beta\leq 4\, 10^{-6}({\rm fm}/\hbar)^2$, or $10^{-4}$ GeV$^{-2}$ in units where $\hbar=c=1$ (that will be used in the rest of this paper). Notice that the minimal length could be system-dependent, thus this bound is stricto sensu valid for electrons.

We propose to focus on the following straightforward generalization of the algebra~(\ref{alg1}) to a $N$-body system:
\begin{eqnarray}
	\left[\hat x_j,\hat p_k\right]&=&i\, \delta_{jk}\, (1+\beta\hat p^2_k),\\
	\left[\hat x_j,\hat x_k\right]&=&	\left[\hat p_j,\hat p_k\right]=0,\nonumber	
\end{eqnarray}
where $ j,k=1,\dots,N$. Notice that no summation is meant in the first line: The commutativity between the coordinates of different particles has been kept, as in standard quantum mechanics. Moreover, the inequality~(\ref{eq2}) holds separately for each particle. The $N$-body Hamiltonian we are interested in reads
\begin{equation}\label{ham0}
\hat H^{(N)}=\sum^N_{j=1}\frac{\hat p^2_j}{2m}+\sum^N_{j<k=1}V(\hat x_j-\hat x_k),	
\end{equation}
that is the Hamiltonian describing a one-dimensional system of $N$-particles with mass $m$ interacting via the pairwise potential $V$. The $N$-body problem with a minimal length has been studied in Ref.~\cite{quesne3}, where macrocsopic (classical) systems are considered. In particular, the analysis of Mercury's perihelion precession leads to the upper bound $0.024$ fm for the minimal length for quarks. In the limit of very large $N$, it is worth mentioning that the modifications of statistical physics due to a nonzero value of $\beta$ have also been discussed in~\cite{Fityo:2008zz,Wang:2010ct}. Thus, to our knowledge, no result concerning the quantum $N$-body problem with a minimal length is currently known. 

The present paper is organized as follows. It is first shown in Sec.~\ref{lbsec} that a lower bound for the ground-state energy of Hamiltonian~(\ref{ham0}) can be obtained in terms of the ground state of a corresponding two-body problem. Then, the scaling in $N$ of the $\beta$-dependent corrections is studied in Sec.~\ref{obsec} at first order in $\beta$. Those general results are particularized to the case of a harmonic interaction potential in Sec.~\ref{hosec}. Finally, the results are summarized in Sec.~\ref{conclu}, and comments are given concerning their validity in $D$-dimensions and for generic interaction potentials.

\section{General formalism}\label{genesec}
\subsection{Lower bound}\label{lbsec}

In standard quantum mechanics, it is quite natural to work with the relative positions $r_{jk}=x_j-x_k$ when dealing with systems whose potential is of the form $V(x_j-x_k)$ -- notice that the symbols without hats denote the standard operators used with the unmodified Heisenberg algebra. The relative momenta are then defined as $\pi_{jk}=(p_j-p_k)/2$ so that the Heisenberg algebra $[r_{jk},\pi_{jk}]=i$ is recovered with the relative coordinates also. The shape of Hamiltonian~(\ref{ham0}) therefore suggests to introduce the ``modified" relative positions
 \begin{equation}
\hat r_{jk}=\hat x_j-\hat x_k	,
\end{equation}
and to define by analogy with the standard case the corresponding relative momenta $\hat\pi_{jk}$ such that the modified Heisenberg algebra
\begin{equation}\label{algr}
	\left[\hat r_{jk},\hat \pi_{jk}\right]=i\, (1+\beta\hat \pi_{jk}^2)
\end{equation}
is obtained. The general form of the commutator $\left[\hat r_{jk},\hat \pi_{lm}\right]$ with $j\neq l$ and $k\neq m$ is not needed since the final expressions we will find are actually separable with respect to the relative coordinates. To find the explicit form of $\hat\pi_{jk}$, we recall that, as suggested by Eq.~(\ref{p2}), the modified Heisenberg algebra we consider comes from an expansion in $\hat p$: It is thus sufficient for our formulas to be valid at order $\hat p^2$. One can then check that the commutation relation~(\ref{algr}) in which $\hat \pi_{jk}$ is equal to 
\begin{equation}
 \hat \pi_{jk}=\left(\frac{\hat p_j-\hat p_k}{2}\right)\left(1-\frac{\beta}{4}(\hat p_j+\hat p_k)^2\right)
\end{equation}
is satisfied at the second order in the momenta $\hat p_j$ as required. It is antisymmetric in $j$, $k$ and reduces to the standard relative momentum for $\beta=0$, as expected. 

Still at the second order in the momenta $\hat p_j$, one has
\begin{equation}
\frac{4}{N}\sum^N_{j<k=1}\hat\pi_{jk}^2=\sum_{j=1}^N\hat p^2_j-\frac{1}{N}\left(\sum^N_{j=1}\hat p_j\right)^2\leq \sum_{j=1}^N\hat p^2_j.
\end{equation}
The above inequality  yield the following lower bound of Hamiltonian~(\ref{ham0}):
\begin{equation}\label{lowb}
	\hat H^{(N)}\geq \sum^N_{j<k=1}\left[\frac{\hat\pi^2_{jk}}{2\mu}+V(\hat r_{jk})\right],
\end{equation}
with
\begin{equation}\label{mudef}
	\mu=\frac{mN}{4}.
\end{equation}
Since the lower-bound Hamiltonian~(\ref{lowb}) is separable, it can be shown that a lower bound on the ground-state energy ${\cal E}^{(N)}$ of $\hat H^{(N)}$ is given by~\cite{kinetic} 
\begin{equation}\label{lowb2}
{\cal E}^{(N)}\geq E^{(N)}=\frac{N(N-1)}{2}\, E^{(2)},
\end{equation}
where $E^{(2)}$ is the ground-state energy of the two-body Hamiltonian
\begin{equation}\label{H2}
\hat	H^{(2)}=\frac{\hat\pi^2}{2\mu}+V(\hat r).
\end{equation}
Remark that in $\hat	H^{(2)}$, $\hat r$ and $\hat \pi$ satisfy $	\left[\hat r,\hat \pi\right]=i\, (1+\beta\hat \pi^2)$. The lower bound~(\ref{lowb2}) implicitly assumes that the spatial wave function of the bound state is totally symmetric. It is consequently valid either for bosons or for fermions when the extra degrees of freedom (spin, isospin, color,\dots) bring an antisymmetric wave function allowing to symmetrize the ground state also. 

Following Eq.~(\ref{lowb2}), any two-body problem with modified Heisenberg algebra whose ground-state energy is known can be used to bound from below the ground-state energy of a corresponding $N$-body problem. It should be stressed that the mass $\mu$ appearing in $H^{(2)}$ is proportional to $N$ as shown by the definition~(\ref{mudef}). Since the $\beta$-dependent terms have not necessarily the same dependence in $\mu$ as the $\beta$-independent ones, this is a first indication that the corresponding dependences in $N$ might be different also. 

\subsection{$O(\beta)$-approach}\label{obsec}

The smallness of $\beta$ with respect to typical quantum-mechanical energy scales suggest that working at the first order in $\beta$ should be relevant. In that case, it is convenient to work with the representation of Ref.~\cite{Brau1}, that can be generalized to the $N$-body case as follows 
\begin{equation}\label{rep}
	\hat x_j=x_j,\quad\hat p_k=\left(1+\frac{\beta}{3}p^2_k\right)p_k ,
\end{equation}
with the standard position and momentum operators satisfying $\left[x_j,p_k\right]=i\, \delta_{jk}$. Using the representation~(\ref{rep}), the two-body Hamiltonian~(\ref{H2}) reads, at the first order in $\beta$,
\begin{equation}\label{H2b}
\hat H^{(2)}=\frac{\pi^2}{2\mu}+V(r)+\frac{\beta}{3\mu}\pi^4.
\end{equation}
Let us now choose the case of a power-law potential 
\begin{equation}
V(x)=\Lambda\, {\rm sgn}(a)\, |x|^a.
\end{equation}
Applied to the one-dimensional case, the virial theorem~\cite{Brau1,afm} then leads to 
\begin{equation}
\left\langle \pi^2\right\rangle=\frac{2\mu a}{a+2} \, E^{(2)}=\frac{ a}{2(a+2)}N\, m\,  E^{(2)}
\end{equation} 
and thus to
\begin{equation}
\frac{\left\langle \pi^4\right\rangle}{\mu}\propto\mu\, (E^{(2)})^2\propto N m\,(E^{(2)})^2.
\end{equation}

Scaling arguments impose that $E^{(2)}=\Lambda^{\frac{2}{a+2}} \mu^{-\frac{a}{a+2}}\, e(a,n)\equiv \Lambda^{\frac{2}{a+2}} (m\, N)^{-\frac{a}{a+2}}\, e_0(a,n)$, where $e$ and $e_0$ are dimensionless functions of $a$ and of a quantum number $n$~\cite{afm1}. One finally gets from Eq.~(\ref{lowb2}) that the lower bound of the ground-state energy is schematically given at large $N$ by
\begin{equation}\label{eneN}
E^{(N)}\approx N^{\frac{a+4}{a+2}}\, \Lambda^{\frac{2}{a+2}}\, m^{-\frac{a}{a+2}}\, e_0\, \left[1+\beta\, (m\Lambda\, N)^{\frac{2}{a+2}}e_1\right],
\end{equation}
where $e_0$ and $e_1$ are dimensionless functions of $a$.
This last relation suggests that the $\beta$-dependent term of the ground-state energy increases with $N$ faster than the $\beta$-independent one, whose dependence in $N$ agrees with the recent analytical calculation~\cite{afmN}. The more $a$ is small (especially when $a$ is negative), the more this effect is significant. For the Coulomb case for example, one would have schematically at large $N$: $E^{(N\gg1)}\sim m N^3+\beta\, m^3N^5$.   

\section{The harmonic oscillator}\label{hosec}

Let us now particularize the results obtained so far to a $N$-body harmonic oscillator, \textit{i.e.} to a potential of the form
\begin{equation}
	V(\hat x_i-\hat x_j)=\Omega\, (\hat x_i-\hat x_j)^2.
\end{equation}
In that case, the ground-state energy can also be easily bounded from above. Indeed, 
\begin{eqnarray}
	\hat H^{(N)}&=&\sum^N_{j=1}\left[\frac{\hat p^2_j}{2m}+\Omega N \hat x^2_j\right]-\Omega \left(\sum^N_{j=1}\hat x_j\right)^2,\nonumber\\
	&\leq&
	\sum^N_{j=1}\left[\frac{\hat p^2_j}{2m}+\Omega N \hat x^2_j\right],
\end{eqnarray}
and the upper bound of the ground-state energy reads
\begin{equation}\label{upb}
{\cal E}^{(N)}\leq N\, \left\langle \frac{\hat p^2}{2m}+\Omega N \hat x^2\right\rangle,
\end{equation}
where $[\hat x,\hat p]=i(1+\beta \hat p^2)$ and where the average is computed with the ground-state wave function. 

The exact spectrum of the two-body harmonic oscillator has been exactly computed in Refs.~\cite{k0b,minic}. It can be deduced from those results that 
\begin{equation}
\left\langle \frac{\hat p^2}{2\nu}+\theta\, \hat x^2\right\rangle=\sqrt{\frac{\theta}{2\nu}+\frac{\beta^2\theta^2}{4}}+\frac{\beta\theta}{2}.
\end{equation}
Combining this last result to the lower and upper bounds~(\ref{lowb2}) and (\ref{upb}) leads to the conclusion that the ground-state energy of a one-dimensional $N$-body harmonic oscillator is bounded by
\begin{eqnarray}
	(N-1)&&\left[\sqrt{\frac{\Omega N}{2m}+\left(\frac{\beta\Omega N}{4}\right)^2}+\frac{\beta\Omega N}{4}\right]\leq {\cal E}^{(N)}
	\leq N \left[\sqrt{\frac{\Omega N}{2m}+\left(\frac{\beta\Omega N}{2}\right)^2}+\frac{\beta\Omega N}{2}\right].
\end{eqnarray}
The lower bound is actually exact for $N=2$ and arbitrary $\beta$, as well as for $\beta=0$ and arbitrary $N$. Furthermore the upper bound is exact for $N=1$ and arbitrary $\beta$. At the first order in $\beta$, the above inequalities become
\begin{equation}
	(N-1)\sqrt{\frac{\Omega N}{2m}} +\frac{\beta \Omega N(N-1)}{4} \leq {\cal  E}^{(N)}\leq N\sqrt{\frac{\Omega N}{2m}} +\frac{\beta \Omega N^2}{2}. 
\end{equation}
At large $N$, the exact ground-state energy is thus of the form 
\begin{equation}
	{\cal E}^{(N\gg1)}=	\sqrt{\frac{\Omega}{2m}}\, N^{\frac{3}{2}} + \beta\, A\, \Omega N^2,
\end{equation}
where $A\in[1/2,1/4]$ is a coefficient independent of $m$. This formula is in qualitative agreement with the lower bound estimate~(\ref{eneN}) for $a=2$, that seems thus to provide a reliable estimation of the behavior in $N$ of the exact ground-state energy.

\section{Summary and discussion}\label{conclu}

The one-dimensional quantum $N$-body problem has been studied within the framework of a modified Heisenberg algebra leading to a minimal length. The system under study is made of $N$ nonrelativistic particles with a mass $m$, interacting via a pairwise potential. We have shown that the ground-state energy can be bounded from below by a convenient formula that only requires to know the ground-state energy of a corresponding two-body system. It is also possible to work at the first order in $\beta$; the correction term is then simply related to the averaged fourth power of the two-body relative momentum. The formalism developed has been explicitly applied to the case of harmonic interactions to check that the several formulas obtained are coherent with each other once applied to a common case. In particular, we can conclude that the ground-state energy of the one-dimensional $N$-body harmonic oscillator is of the form $b N^{\frac{3}{2}} + \beta c N^2$ at large $N$. As illustrated by this last relation, it appears from the present study that for power-law potentials, the $\beta$-dependent term grows faster with $N$ than the $\beta$-independent energy, the effect being especially important for singular attractive potentials of the form $-1/|x|^k$. 

Somme comments can be done concerning the extension of the present results to more realistic potentials and higher-dimensional systems. First, provided that it is not located too close of the continuum, the ground-state will be mostly sensitive to the short-range behavior of the potential $V(x)$, that can be approximated by its Taylor expansion near $x=0$. One can thus say that, in a first approximation, the faster increase with $N$ of the $\beta$-dependent term, obtained for power-law potentials, will qualitatively be a feature of generic potentials if the ground-state binding energy is significant. Second, $D$-dimensional generalizations of the modified $N$-body algebra we considered generally depend on two parameters denoted $\beta$ and $\beta'$~\cite{quesne3}. In analogy with Sec.~\ref{lbsec}, let us assume that we know the relative coordinates $\hat{\bm r}_{jk}=\hat{\bm x}_j-\hat{\bm x}_k$, $\hat{\bm\pi}_{jk}=(\hat{\bm p}_j-\hat{\bm p}_k)(1+s(\beta,\beta',\hat{\bm p}_j,\hat{\bm p}_k))/2$, where the bold symbols denote vectors, such that the relative coordinates satisfy the same algebra as the particles coordinates. The function $s$ should be such that $s(0,0,\hat{\bm p}_j,\hat{\bm p}_k)=0$ and $s(\beta,\beta',\hat{\bm p}_j,\hat{\bm p}_k)=s(\beta,\beta',\hat{\bm p}_k,\hat{\bm p}_j)$. Assuming that it has a nontrivial dependence in the momenta, one would find again the lower bound~(\ref{lowb2})-(\ref{H2}) at second order in the momenta. A case of interest is the algebra in which $\beta'=2\beta$, keeping commutative positions at $O(\beta)$, and admitting at this order the representation $\hat{\bm x}_j={\bm x}_j$, $\hat{\bm p}_k=(1+\beta {\bm p}_k^2){\bm p}_k$~\cite{Brau1}. One would finally find the term $\beta\, {\bm\pi}^4/\mu$ in the $D$-dimensional generalization of~(\ref{H2b}), leading again to the estimation~(\ref{eneN}) for the ground-state energy in the case of radial power-law potentials. 

In conclusion, it has been shown that there exists a nontrivial interplay between the quantum $N$-body dynamics and the existence of a minimal length, whose manifestation is an enhancement of the minimal length effects at large number of particles. Although explicit examples are out of the scope of the present work, this suggests that the comparison between high-precision models and measurements related to quantum $N$-body systems (atomic, molecular, etc.) might eventually be an interesting and new way of constraining the value of a minimal length in Nature.

\acknowledgments
I thank the F.R.S.-FNRS for financial support, F. Brau for having drawn my attention on the $N$-body problem with a minimal length, and C. Semay for valuable comments about the present manuscript.

\end{document}